\documentclass[intlimits,twoside,a4paper]{article}

\usepackage{graphicx}        
\usepackage[cp1251]{inputenc}

\usepackage[eqsecnum]{cmpj3}

\issue{2022}{25}{4}{43702}
\doinumber{10.5488/CMP.25.43702}

\title[Ferroelectric instability of barium sodium niobate]%
{Broadband Brillouin scattering study of ferroelectric instability of barium sodium niobate}
\author[S. Kojima]{S. Kojima\thanks{e-mail: {kojima@ims.tsukuba.ac.jp}}}
\address{Division of Materials Science, University of Tsukuba, Tsukuba, Ibaraki 305-8573, Japan}
\date{Received July 18, 2022}

\begin{document}

\maketitle

\begin{abstract}
The barium sodium niobate (BNN) with tungsten-bronze structure is one of 
well-known optical crystals for electro-optic and nonlinear optic 
applications. This paper reviews the ferroelectric instability of BNN 
crystals. BNN is a uniaxial ferroelectric with a spontaneous polarization 
along the tetragonal $c$-axis. There is no report on the observation of an 
optical soft mode responsible for a ferroelectric phase transition. In the 
vicinity of the Curie temperature, $T_{\text{C}}=560$\textcelsius{}, an intense central 
peak (CP) related to the polarization fluctuations along the $c$-axis was 
observed by the broadband Brillouin scattering experiment. The relaxation 
time determined by the~CP width shows the critical slowing down towards 
$T_{\text{C}}$. This fact indicates that the ferroelectric instability of BNN is an 
order-disorder type.
\end{abstract}

\section{Introduction}

Ferroelectricity is defined by the existence of switchable spontaneous 
polarization by an external electric field. Ferroelectric phenomenon was 
identified for the first time in 1920 by Valasek on the study of Rochelle 
salt~\cite{Kojima_ref1}. The microscopic origin of ferroelectricity has two typical cases, 
namely, the displacive type and order-disorder type. In the displacive type, 
an infrared active soft optic mode exists in a paraelectric phase, and the 
freezing of a soft mode displacement induces a spontaneous polarization~\cite{Kojima_ref2,Kojima_ref3}. The softening of a soft mode frequency has been observed by 
far-infrared spectroscopy, Raman scattering, and neutron inelastic 
scattering. On the other hand, in the order-disorder type, the relaxation 
time of the polarization fluctuations of polar molecules along a 
ferroelectric axis diverges at the Curie temperature, and the aligned polar 
molecules induce a spontaneous polarization. Rochelle salt belongs to this 
type. The critical slowing down towards a Curie temperature has been 
observed by dielectric spectroscopy. In NaNO$_{2}$, Hatta observed the 
divergence of the relaxation time of the flipping motion of each NO$_{2}$ 
ion toward a Curie temperature due to the thermodynamical slowing down 
process of the correlated fluctuation of polarization~\cite{Kojima_ref4}. The critical 
slowing-down of the polarization relaxation process was also observed in 
triglycine sulfate~\cite{Kojima_ref5} and Ca$_{2}$Sr(C$_{2}$H$_{5}$CO$_{2})_{6}$ above 
the Curie temperature~\cite{Kojima_ref6}.  Another method to observe the critical 
slowing down is the low-frequency inelastic light scattering. The 
polarization fluctuations along a ferroelectric axis are observed as a 
central peak (CP). In the vicinity of a Curie temperature, the divergence of CP 
intensity and the narrowing of CP width are observed for an order-disorder 
phase transition. In a K(Ta$_{0.68}$Nb$_{0.32})$O$_{3}$ crystal, the 
relaxation time determined by the CP width clearly shows a critical slowing 
down towards the Curie temperature, $T_{\text{C}}=258$~K, indicating an 
order-disorder feature of the ferroelectric phase transition~\cite{Kojima_ref7,Kojima_ref8}. Up to 
the present, the critical slowing down has been studied by the observation 
of a CP in ferroelectric phase transitions of 12 mol{\%} KF 
substituted BaTiO$_{3}$~\cite{Kojima_ref9}, LiTaO$_{3}$~\cite{Kojima_ref10}, KNN~\cite{Kojima_ref11}, MAPbCl$_{3}$~\cite{Kojima_ref12}, 
BaTi$_{2}$O$_{5}$~\cite{Kojima_ref13}, and K$_{2}$MgWO$_{2}$(PO$_{4})_{2}$~\cite{Kojima_ref14}. 

For the ferroelectric phase transitions of relaxor ferroelectrics, the 
diffusive nature was observed in the critical slowing down~\cite{Kojima_ref15}. In 
0.70Pb(Zn$_{1/3}$Nb$_{2/3})$O$_{3}$--0.30PbTiO$_{3}$ (PZN--7PT), the slowing 
down was suppressed below the intermediate temperature $T^{*}$ and the typical 
critical slowing down was not observed near $T_{\text{C}}$. The local transition 
from dynamic to static PNRs at $T^{*}$ stops the farther slowing down. To 
describe such a suppressed slowing down by random fields, the empirical 
equation of the stretched slowing down was used in the vicinity of $T_{\text{C}}$ as 
given by the following equation,
\begin{equation}
	\frac{1}{\tau_{\text{CP}}} 
	= 
	\frac{1}{\tau_{0}} 
	+ 
	\frac{1}{\tau_{1}}
		\left(\frac{T - T_{\text{C}}}{T_{\text{C}}}\right)^{\beta}
	,
	\quad
	(T>T_{\text{C}}),
\label{Kojima_eq1}
\end{equation}
where $\beta$ is the stretched index. In the case of $\beta=1.0$, the equation~\eqref{Kojima_eq1} gives a critical slowing down of normal ferroelectrics without random fields. In the case of $\beta>1.0$, the slowing down of relaxation time is suppressed and/or stretched by an increase of the strength of random fields. In PZN--7PT, it is found that the value of $\beta=3.0$ gives a good reproduction of slowing down~\cite{Kojima_ref16}. 

In some ferroelectrics, the mechanism of the ferroelectricity is not simple. 
The coexistence of both mechanisms or the crossover from displacive to 
order-disorder nature has been reported. In LiNbO$_{3}$, two-stage process 
involving a displacive transition in the Nb--O cages and an order-disorder 
transition in the Li--O planes was reported at the Curie temperature~\cite{Kojima_ref17}. 
The unified model theory describing both the ``order-disorder'' and 
``displacive'' ferroelectric phase transitions was proposed by introducing the 
model pseudospin-phonon Hamiltonian~\cite{Kojima_ref18}. For such phase transitions, the 
study of the lowest frequency soft optic modes by Raman scattering or 
infrared spectroscopy is also necessary.

\section{Ferroelectrics with tungsten-bronze structure}

Ferroelectricity has been observed in various kinds of organic and inorganic 
materials. Regarding inorganic ferroelectrics, the oxygen octahedra 
ferroelectrics are the most popular. One of this family is the 
ferroelectrics with tungsten-bronze structure. It is technologically 
important in the field of telecommunications due to its superior 
electro-optical, photorefractive, and nonlinear optical properties such as 
second harmonic generation, and its resistance to optical damage is high. 
The ferroelectricity was reported at first in lead metaniobate, 
PbNb$_{2}$O$_{6}$ with $T_{\text{C}}=570$\textcelsius{}~\cite{Kojima_ref19}. Its piezoelectric constant is the same order of magnitude as that of barium titanate, BaTiO$_{3}$ with
$T_{\text{C}}=120$\textcelsius{}. The useful nonlinear coefficients and low optical damage were 
reported in barium sodium niobate, Ba$_{2}$NaNb$_{5}$O$_{15}$ (BNN), with 
$T_{\text{C}}=560$\textcelsius{}~\cite{Kojima_ref20}. Nowadays, a lot of tungsten-bronze type 
ferroelectrics are known~\cite{Kojima_ref2}.

\begin{figure}[!b]
\centerline{\raisebox{-0.5\height}{\includegraphics[width=0.35\textwidth]{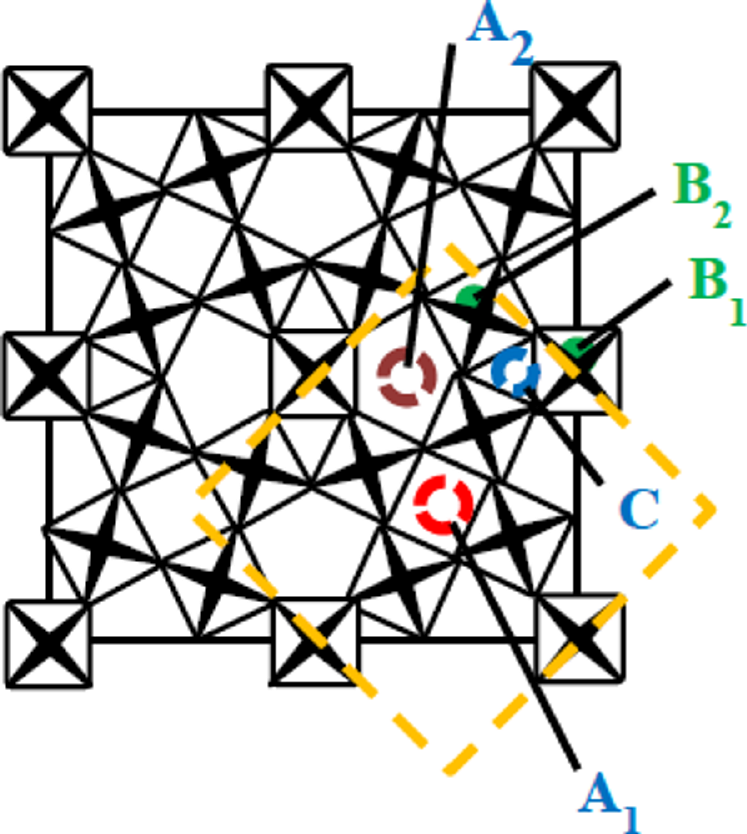}}%
\qquad%
\raisebox{-0.5\height}{\includegraphics[width=0.25\textwidth]{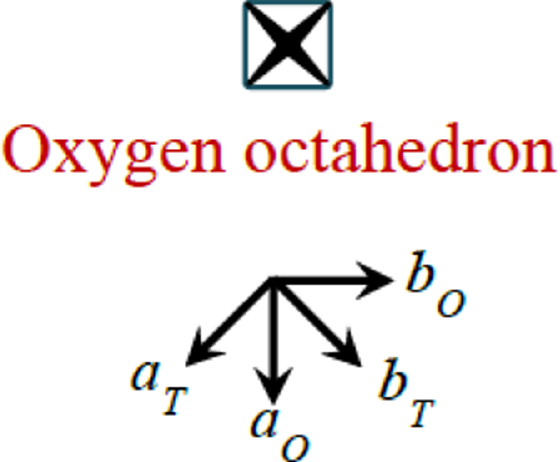}}}
%
%
\caption{(Colour online) Projection of tungsten bronze structure on the $c$-plane. Tetragonal 
and orthorhombic unit cells are shown by dotted and solid lines, 
respectively.}
\label{Kojima_fig1}
\end{figure}

Figure~\ref{Kojima_fig1} shows the projection of tungsten bronze structure on the $c$-plane. In 
BNN, the A$_{1}$ and A$_{2}$ sites are fully filled by Na and Ba, 
respectively, and there is no vacancy. The B$_{1}$ and B$_{2}$ sites are 
fully occupied by Nb, while the C sites are vacant. BNN is called a filled 
tungsten-bronze structure with no charge disorder at A$_{1}$ and A$_{2}$ 
sites. BNN undergoes successive phase transitions at 560, 300, and $-163$\textcelsius{}~\cite{Kojima_ref21}. The higher one is associated with the ferroelectric Curie 
temperature at 560\textcelsius{}, and its crystal symmetry changes from prototypic 
$4/mmm$ to ferroelectric tetragonal $4mm$ with a spontaneous polarization $P_{3}$ along 
the $c$-axis. The symmetry changes from tetragonal to incommensurate (IC) 
orthorhombic $mmm$ systems at 300\textcelsius{}, the $a$ and $b$ axes are rotated for 
45$^{\circ}$ along the $c$-axis as shown in figure~\ref{Kojima_fig1}. The modulation direction of the IC wave vectors is along $a$ and $b$ axes of the orthorhombic coordinate~\cite{Kojima_ref22}. The lowest temperature phase transition at $-163$\textcelsius{} is the reentrant ferroelastic phase transition into the tetragonal $4mm$ phase. The pressure 
induced reentrant ferroelastic phase transition was also observed at~2.2~GPa 
and at room temperature~\cite{Kojima_ref23,Kojima_ref24}. Recently, a new type of the IC phase 
transition was proposed by Ishibashi~\cite{Kojima_ref25}. This new type of phase transition 
is referred to as type~III, and it is characterized by the parabolic splitting 
of the doubly degenerate modes at the Brillouin zone boundary. The related 
macroscopic change in the IC phase transition was studied by Brillouin 
scattering~\cite{Kojima_ref26}. Regarding a ferroelectric instability, the accurate 
measurement of low-frequency polaritons was performed on the optical phonon 
branch of $\mathrm{A}_{1}(z)$ symmetry. However, any evidence of displacive nature 
was not found down to 18~cm$^{-1}$~\cite{Kojima_ref27}.

Since the $T_{\text{C}}$ of Ba$_{2}$NaTa$_{5}$O$_{15}$ (BNT) is $-233$\textcelsius{}, the 
high tunability of the $T_{\text{C}}$ of the temperature width of about 800\textcelsius{} 
was reported for Ba$_{2}$NaNb$_{5(1-x)}$Ta$_{5x}$O$_{15}$ (BNNT), and this is 
technologically important~\cite{Kojima_ref28,Kojima_ref29}. The ferroelectric phase transition of the 
BNNT single crystals with $x=0.57$ at $T_{\text{C}}=115$\textcelsius{} was studied by 
broadband dielectric spectroscopy up to 4~GHz. The order-disorder nature of 
the proper ferroelectric phase transition was observed, and its origin is 
attributed to the anharmonic motion of the Nb (Ta) atoms in a double well 
potential of oxygen octahedra~\cite{Kojima_ref30}. 

Up to the present, no soft optic mode was observed in the tungsten-bronze 
ferroelectrics by vibrational spectroscopy. Recent theoretical studies 
reported the local pseudo-Jahn--Teller effect (PJTE) in transition metal B 
ion center of ABO$_{3}$ perovskite crystals. The vibronic coupling 
between the ground and excited electronic states of the local BO$_{6}$ 
center results in dipolar distortions, leading to an eight-well adiabatic 
potential energy surface~\cite{Kojima_ref31}. Such a situation may also occur in tungsten 
bronze ferroelectrics and the order-disorder nature of ferroelectricity may 
exist. Therefore, the order-disorder nature of a ferroelectric phase 
transition of BNN has been examined by the broadband Brillouin scattering 
spectroscopy, which is a powerful tool to observe a critical slowing down in 
the vicinity of an order-disorder type phase transition temperature. In this 
paper, we review the broadband Brillouin scattering study on the 
ferroelectric instability of a BNN crystal~\cite{Kojima_ref32}.

\section{Broadband Brillouin scattering and ferroelectric instability}

Vibrational spectroscopy i.e., infrared spectroscopy and Raman scattering 
observes the vibrational modes of atoms, molecules, and crystal lattice. In 
the vibrational study of inorganic ferroelectric crystals, it is possible to 
observe not only the internal modes of octahedra or tetrahedra but also the 
external modes such as a soft optic mode. The spectral resolution of 
vibrational spectroscopy is usually 1~cm$^{-1}$${}=30$~GHz or more, and it is 
sufficient to detect the change of mode frequency related to a phase transition. 
The resolution of 1~cm$^{-1}$ is sufficient to measure the temperature 
dependence of a ferroelectric soft optic mode. However, in the study of a 
ferroelectric phase transition of order-disorder type, the resolution of~1~cm$^{-1}$ is not sufficient to measure the critical slowing down of the 
relaxation time towards the Curie temperature~\cite{Kojima_ref33}.

\begin{figure}
\centerline{\includegraphics[width=0.65\textwidth]{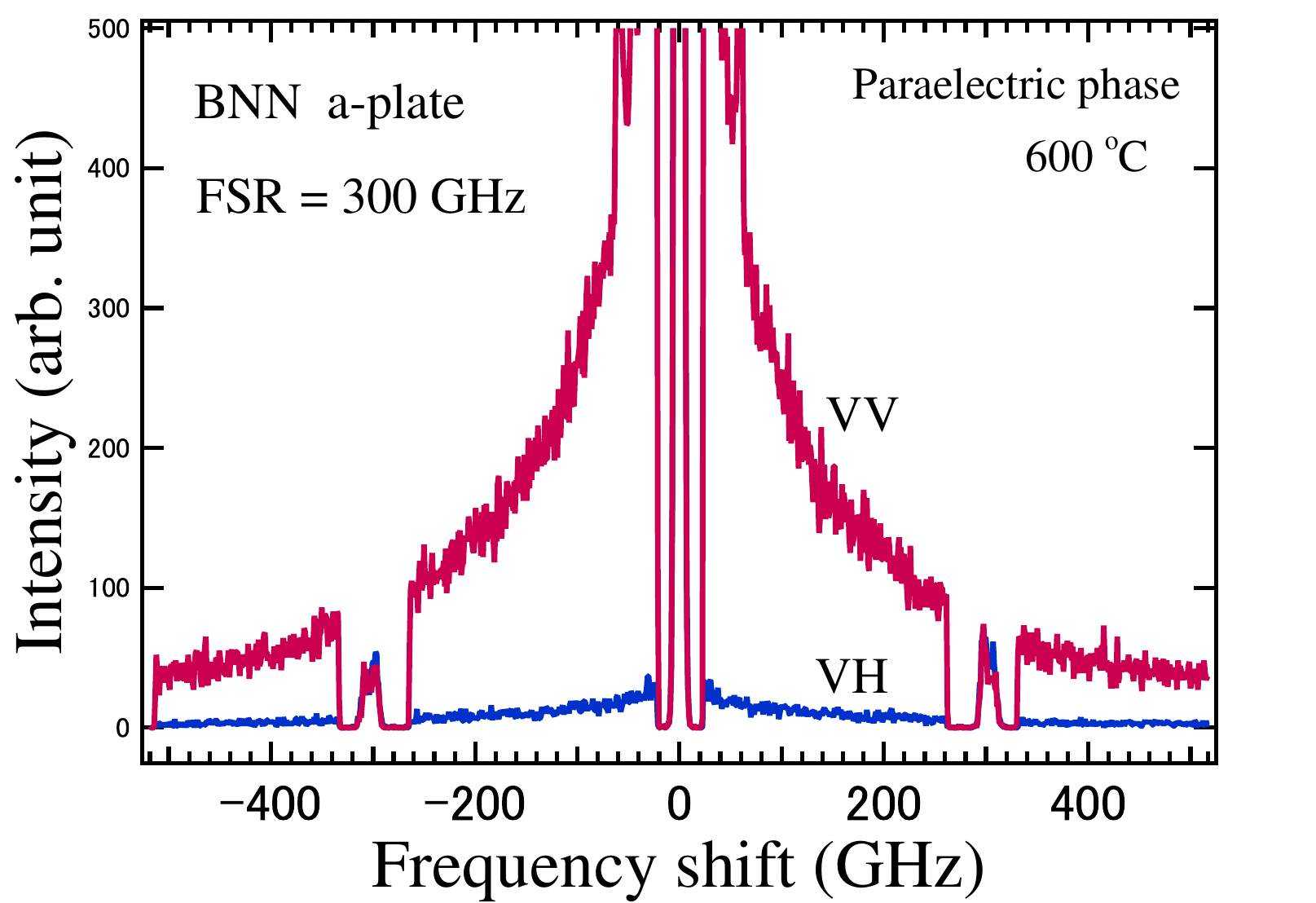}}%
\caption{(Colour online) Broadband VV and VH Brillouin scattering spectra of a BNN crystal at 600\textcelsius{} observed by $a(cc)\bar{a}$ and $a(cb)\bar{a}$ back scattering geometry, respectively.}
\label{Kojima_fig2}
\end{figure}

Polarization fluctuations related to a ferroelectric instability are 
detected as a broad CP in an inelastic scattering spectrum. The 
colorless and transparent BNN crystal studied was grown by Czochralski 
method in Tamagawa factory, NEC, Japan. The (100) plate with the size of 
$2.65 \times 2.25 \times 0.68$~mm with two optically polished surfaces was 
used for Brillouin scattering measurements. Brillouin scattering spectra 
were measured at the backward scattering geometry using a $3+3$ tandem 
multi-pass Fabry--Perot interferometer and a conventional photon counting 
system. A single frequency green YAG laser ($\lambda=532$~nm) with power 
of 100~mW was used as an exciting source. The light spot size at a sample 
surface was about 10~$\mu$m using the optical microscope (BX-60)~\cite{Kojima_ref33}. The 
temperature of a sample was controlled by the heating stage of a T1500 (high 
T Linkam) from room temperature up to 750\textcelsius{}. All the Brillouin 
scattering spectra were measured in the condition that the free spectral 
range (FSR) and the scan range are 300 and 600~GHz, respectively. An intense 
polarized CP of BNN was observed in the vicinity of the Curie 
temperature, $T_{\text{C}}=560$\textcelsius{}, in the broadband Brillouin scattering 
spectra as shown in figure~\ref{Kojima_fig2}~\cite{Kojima_ref34}. In the polarized VV spectrum observed at $a(cc)\bar{a}$, backward scattering geometry shows an intense broad CP with 
$\mathrm{A}_{1}(z)$ symmetry, while in the depolarized VH spectrum observed at 
$a(cb)\bar{a}$, backward scattering geometry does not show an intense CP 
with $\mathrm{B}_{2}$ symmetry. Therefore, the polarization fluctuations along a ferroelectric $c$-axis are the origin of an intense broad~CP.

\section{Critical slowing down on a ferroelectric phase transition of barium sodium niobate}

\begin{figure}[!h]
\centerline{\includegraphics[width=0.65\textwidth]{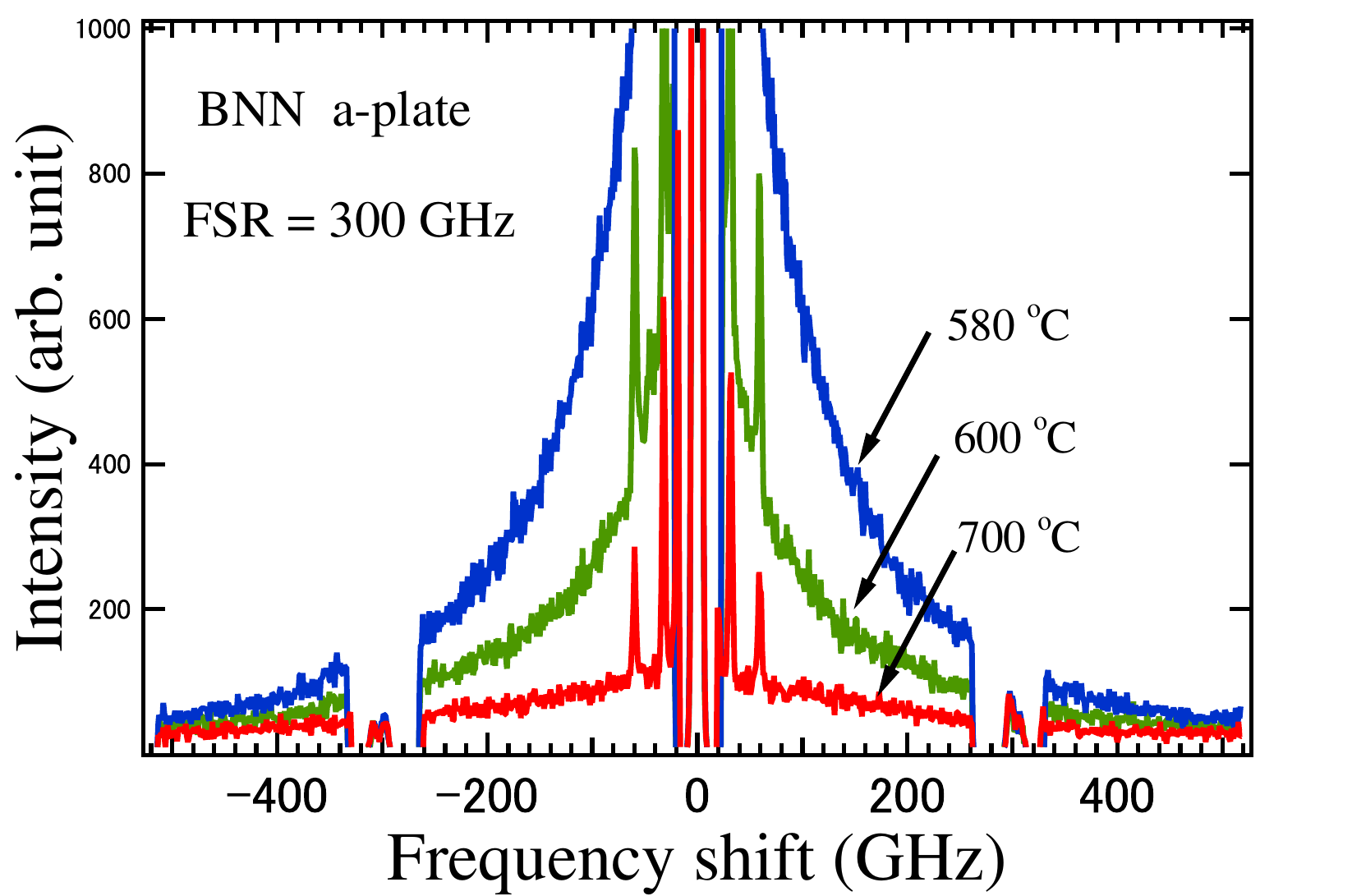}}%
\caption{(Colour online) Broadband VV Brillouin scattering spectra of a BNN crystal in a paraelectric phase.}
\label{Kojima_fig3}
\end{figure}

For the detailed analysis of the width of a CP, the 
temperature dependence of broadband Brillouin scattering spectra of a BNN 
crystal was measured at the backward scattering geometry with the free 
spectral range of 300~GHz as shown in figure~\ref{Kojima_fig3}. 

\begin{figure}[h]
\centerline{\includegraphics[width=0.65\textwidth]{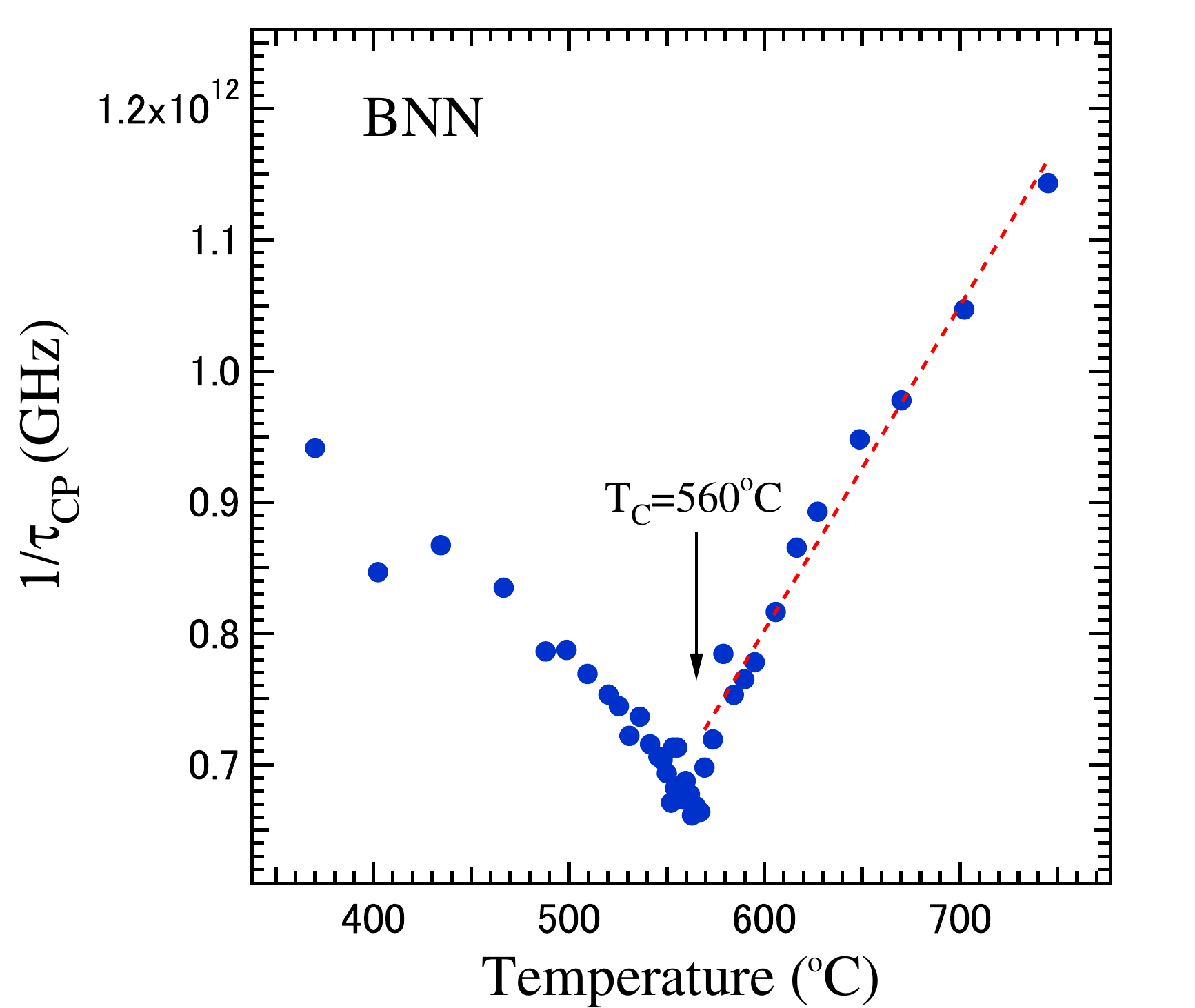}}%
\caption{(Colour online) Temperature dependence of the inverse relaxation time. The dotted line is the fitted line by the equation~\protect\eqref{Kojima_eq2} above $T_{\text{C}}=560$\textcelsius{}.}
\label{Kojima_fig4}
\end{figure}

\begin{figure}[h]
\centerline{\includegraphics[width=0.65\textwidth]{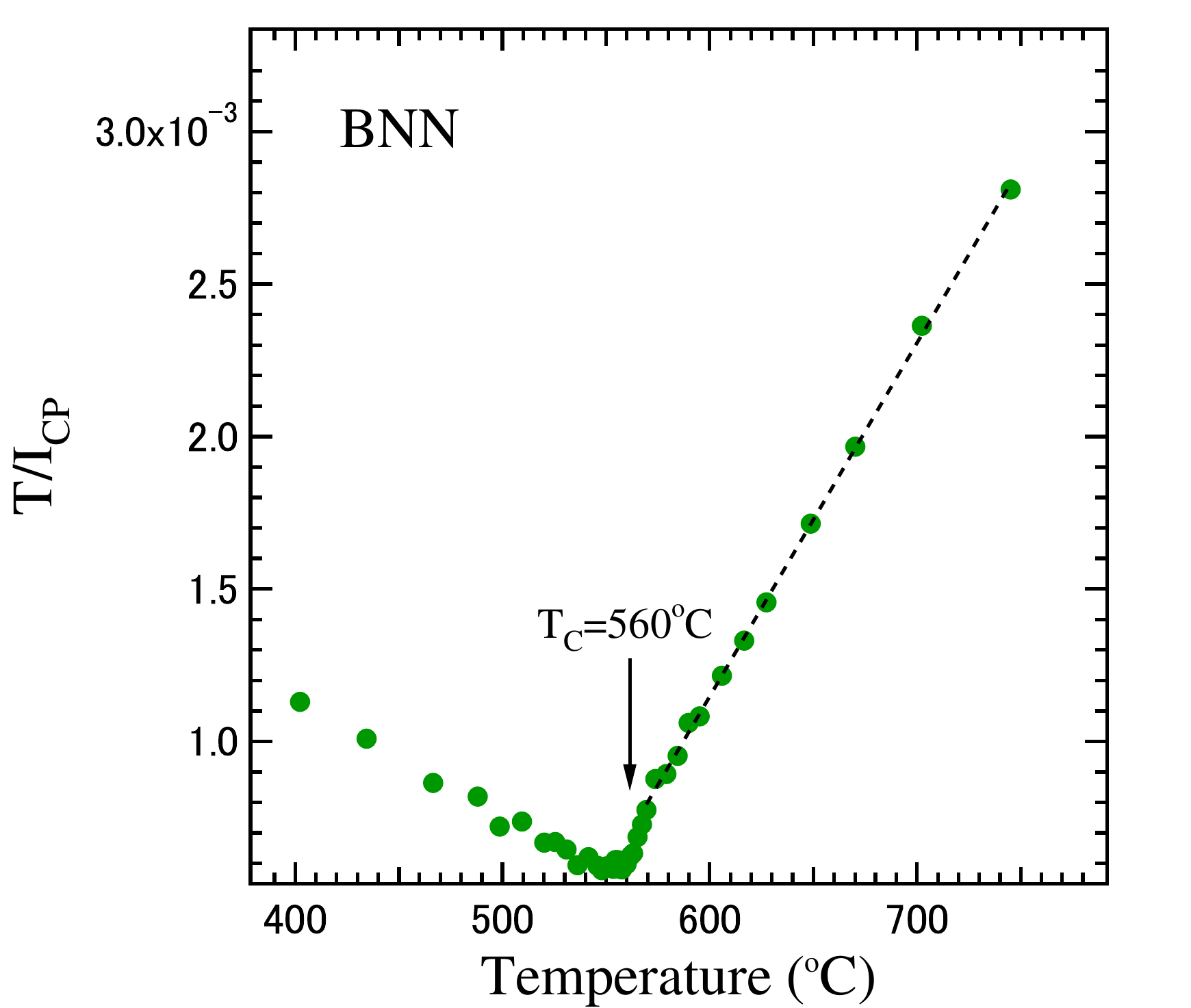}}%
\caption{(Colour online) Temperature dependence of temperature divided by the intensity of a central peak. The dotted line is the fitted line by the equation~\protect\eqref{Kojima_eq3}
above $T_{\text{C}}=560$\textcelsius{}.}
\label{Kojima_fig5}
\end{figure}

Under the assumption of a single Debye relaxation process, the relaxation 
time $\tau_{\text{CP}}$ was determined by the relation 
$\piup \times \text{(CP width)} = \tau_{\text{CP}}^{-1}$. 
The relaxation process related to the order-disorder nature 
of a ferroelectric phase transition has been observed as a CP with a zero 
frequency shift in an inelastic scattering spectrum. In the order-disorder 
phase transition, the relaxation time $\tau$ of the fluctuations of the order 
parameters increases toward the phase transition point and was called the 
critical slowing down. The relaxation time determined from the CP width 
shows a critical slowing down in the vicinity of $T_{\text{C}}=560$\textcelsius{} as 
shown in figure~\ref{Kojima_fig4}~\cite{Kojima_ref34}. The temperature dependence of the relaxation time 
is given by the following equation of the case of $\beta=1.0$ in equation~\eqref{Kojima_eq1} 
for a first order phase transition:
\begin{equation}
	\frac{1}{\tau_{\text{CP}}} 
	= 
	\frac{1}{\tau_{0}} 
	+ 
	\frac{1}{\tau_{1}}
		\left(\frac{T - T_{1}}{T_{1}}\right)
	,
	\quad
	(T>T_{\text{C}}>T_{1})
	.
\label{Kojima_eq2}
\end{equation}
For example, in the ferroelectric phase transition at $T_{\text{C}}=500$~K of the 
relaxor ferroelectric
\linebreak
0.70Pb(Sc$_{1/2}$Nb$_{1/2})$O$_{3}$--0.30PbTiO$_{3}$ 
with the perovskite structure, the values of the fitting parameters are 
$\tau_{0}=14$~ps and $\tau_{1}=0.47$~ps, and $T_{1}=500$~K~\cite{Kojima_ref35}. The temperature 
dependences of $T/I_{\text{CP}}$ of a BNN crystal are shown in figure~\ref{Kojima_fig5}. In BNN, the 
fitting parameters of $1/\tau$ are $\tau_{0}=1.29$~ps, $\tau_{1}=0.73$~ps, 
and $T_{1}=555$\textcelsius{}. The intensity of a CP $I_{\text{CP}}$ obeys the 
following equation in a paraelectric phase~\cite{Kojima_ref36}:
\begin{equation}
	\frac{T}{I_{\text{CP}}} 
	\propto 
	\left[ 
		\int_{0}^{\infty}
		{\frac{\chi''
			\left(\omega\right)}{\omega}\,\mathrm{d}\omega} 
	\right]^{-1} 
	\propto 
	\frac{1}{\chi'(0)} 
	= 
	\frac{T - T_{1}}{C}
	,
	\quad
	(T>T_{\text{C}}>T_{1}).
\label{Kojima_eq3}
\end{equation}
Here, for the first order phase transition, $T_{\text{C}}>$T$_{1}$, because the 
ferroelectric phase transition of BNN is the first order. In the 
ferroelectric phase transition at $T_{\text{C}}=500$~K of the 
0.70Pb(Sc$_{1/2}$Nb$_{1/2})$O$_{3}$--0.30PbTiO$_{3}$, the Curie--Weiss law 
also holds for $I_{\text{CP}}/T$ above $T_{\text{C}}$~\cite{Kojima_ref35}.

The experimental results of the critical slowing down of relaxation time and 
the Curie--Weiss behavior of the CP intensity indicate the order-disorder 
nature of a ferroelectric phase transition of BNN. In the study of the 
order-disorder phase transition, Brillouin scattering is a powerful tool to 
detect the critical slowing down~\cite{Kojima_ref37}.

\section{Conclusions}

For the study of the lattice instability of ferroelectrics, vibrational 
spectroscopy is a powerful tool to discuss not only displacive but also 
order-disorder nature. This paper reviews the experimental studies on the 
ferroelectric instability of a ferroelectric phase transition of barium 
sodium niobate (BNN) crystals with tungsten-bronze structure. BNN is one of 
well-known optical crystals for electro-optic and nonlinear optic 
applications. It shows a uniaxial ferroelectricity with a spontaneous 
polarization along the tetragonal $c$-axis. In the vicinity of the Curie 
temperature, $T_{\text{C}}=560$\textcelsius{}, an intense central peak (CP) was observed 
by the broadband Brillouin scattering experiment. The CP has a strong 
polarization dependence, which originates from the polarization 
fluctuations along the ferroelectric $c$-axis. The CP intensity shows a 
maximum at $T_{\text{C}}$. The relaxation time determined by the CP width shows a 
critical slowing down towards~$T_{\text{C}}$. The temperature dependence of the CP 
intensity shows the Curie--Weiss behavior. These experimental results are the 
evidence of the order-disorder nature of the ferroelectric instability of 
BNN.

\section*{Acknowledgements}

Author thanks to Prof.~J.~Grigas, Prof.~J.~Banys, Prof.~M.~Maczka for the collaboration and S.~Ohta, Y.~Christy, K.~Matsumoto, K.~Suzuki, and M.~Aftebuzzamann for the discussion and experiments.

\section*{Funding}

This research was funded in part by JSPS KAKENHI, Grant No.~JP17K05030.

\ukrainianpart

\title{Дослідження сегнетоелектричної нестійкості в	ніобаті барію-натрію методами
	широкосмугового розсіювання Бріллюена}
\author{С. Коджіма}
\address{Відділення матеріалознавчих наук, Університет Цукуби, Цукуба, Ібаракі 305-8573, Японія}

\makeukrtitle

\begin{abstract}
	\tolerance=3000%
	Ніобат барію-натрію (BNN) зі структурою вольфрамової бронзи є одним з добре відомих оптичних кристалів,
	які використовуються для електрооптичних досліджень та у нелійнійній оптиці. У даній роботі розглядається сегнетоелектрична нестійкість в кристалах BNN. 
	BNN є одновісним сегнетоелектриком, в якому спонтанна поляризація напрямлена вздовж тетрагональної осі $c$. 
	У літературі немає згадок про спостереження оптичної м'якої моди, відповідальної за сегнетоелектричний фазовий перехід у цьому кристалі. В околі температури Кюрі $T_{\text{C}}=560$\textcelsius{} 
	в спектрах широкосмугового розсіювання Бріллюена спостерігається інтенсивний центральний пік, 
	пов'язаний з флуктуаціями поляризації удовж осі $c$. Час релаксації, який визначається шириною центрального піка, виявляє критичне сповільнення при наближенні до
	$T_{\text{C}}$. Цей факт свідчить про те, що сегнетоелектрична нестійкість у BNN-сполуках 
	є типу ``лад-безлад''.
	\keywords розсіювання Бріллюена, сегнетоелектрик, лад-безлад, центральний пік, ніобат барію-натрію
	
\end{abstract}

\lastpage
\end{document}